\documentstyle{article}

\def\noheaderplainsetup{ 

\topmargin=0pt \headheight=0pt \headsep=0pt  \oddsidemargin=0pt \evensidemargin=0pt  \textheight=8.9truein \textwidth=6.5truein}   

\noheaderplainsetup

\newcommand{\chess}{\mbox{\em Chess}}
\newcommand{\father}{\mbox{\em Father}}
\newcommand{\mother}{\mbox{\em Mother}}
\newcommand{\nainai}{\mbox{\em PaternalGrandmother}}
\newcommand{\even}{\mbox{\em Even}}
\newcommand{\odd}{\mbox{\em Odd}}
\newcommand{\intimpl}{\mbox{\hspace{2pt}$\circ$\hspace{-0.14cm} \raisebox{-0.043cm}{\Large --}\hspace{2pt}}}
\newcommand{\brneg}{\mbox{\hspace{2pt}$\circ\hspace{-0.05cm} \neg$\hspace{2pt}}}
\newcommand{\cla}{\mbox{\large $\forall$}}      
\newcommand{\cle}{\mbox{\large $\exists$}}      
\newcommand{\ade}{\mbox{\Large $\sqcup$}}      
\newcommand{\ada}{\mbox{\Large $\sqcap$}}      
\newcommand{\add}{\sqcup}                      
\newcommand{\adc}{\sqcap}                      
\newcommand{\st}{\mbox{\raisebox{-0.05cm}{$\circ$}\hspace{-0.13cm}\raisebox{0.16cm}{\tiny $\mid$}\hspace{2pt}}}
\newcommand{\cost}{\mbox{\raisebox{0.12cm}{$\circ$}\hspace{-0.13cm}\raisebox{0.02cm}{\tiny $\mid$}\hspace{2pt}}}

\newtheorem{theoremm}{Theorem}[section]
\newtheorem{thesiss}[theoremm]{Thesis}

\newenvironment{thesis}{\begin{thesiss} \em}{ \end{thesiss}}
\newenvironment{theorem}{\begin{theoremm}}{\end{theoremm}}

\begin{document}

\title{Computability logic: Giving Caesar what belongs to Caesar}
\author{Giorgi Japaridze
  \\  
 \\ Villanova University  and \\ Institute of Philosophy, Russian Academy of Sciences\\
800 Lancaster Avenue, Villanova, PA 19085, USA \\
 Email: giorgi.japaridze@villanova.edu}

\date{}

\maketitle

\begin{abstract} The present article is a brief informal survey of {\em computability logic} (CoL). This relatively young and still evolving nonclassical logic can be characterized as a formal theory of computability in the same sense as classical logic is a formal theory of truth.  In a broader sense, being conceived semantically rather than proof-theoretically, CoL is not just a particular theory but an ambitious and challenging long-term project for redeveloping logic.  

In CoL, logical operators stand for operations on computational problems, formulas represent such problems, and their ``truth'' is seen as algorithmic solvability. In turn, computational problems --- understood in their most general, interactive sense --- are defined as games played by a machine against its environment, with ``algorithmic solvability''  meaning existence of a machine which wins the game against any possible behavior of the environment. With this semantics, CoL provides a systematic answer to the question ``What can be computed?'', just like classical logic is a systematic tool for telling what is true. Furthermore, as it happens, in positive cases ``What can be computed''  always allows itself to be replaced by ``How can be computed'', which makes CoL a problem-solving  tool.   

CoL is a conservative extension of classical first order logic but is otherwise much more expressive than the latter, opening a wide range of new  application areas. It relates to intuitionistic and linear logics in a similar fashion, which allows us to say that CoL reconciles and unifies the three  traditions of logical thought (and beyond) on the basis of its natural and ``universal'' game semantics.  

\end{abstract}


\ 

The present article is essentially a transcript of a lecture I gave in Moscow at the Institute of Philosophy on October 19, 2017. This explains some peculiarities of its style, such as speaking in first person or absence of formal definitions. Thanks go to my good old friend Vladimir Shalack for organizing the lecture and forcefully nudging me afterwards to write an article based on it. 

\section{Computability logic versus classical logic} Not to be confused with the generic term ``computational logic'', ``computability logic'' (CoL) is the proper name of an approach and ongoing ambitious project initiated by myself back in 2003 \cite{Jap03}.  I characterize it as a ``formal theory of  computability in the same sense as classical logic is a formal theory of  truth''. To see what this means, let us compare the two logics.

\begin{itemize} 
  \item In classical logic, the central semantical concept is {\em truth};  formulas represent {\em statements}; and the main utility of the logic is that it provides a systematic answer to the questions ``What is (always) {\em true}?'' or ``Does {\em truth} of $P$ (always) follow  from {\em truth} of $Q$?''.\footnote{Of course, one is a special case of the other.} 
  \item In computability logic, the central semantical concept  is {\em computability}; formulas represent {\em computational problems}; and the main utility of the logic is that it provides a systematic answer to the questions ``What is (always) {\em computable}?'' or ``Does {\em computability} of $P$ (always) follow  from {\em computability} of $Q$?''. 
\end{itemize}
As we see, the second bulleted item is identical to the first one, only with ``truth'' replaced by ``computability'' everywhere and, correspondingly, ``statements'' by ``computational problems'' (for computability is the desired property of computational problems just like truth is the desired property of statements). In positive cases computability logic additionally provides a systematic answer to 
not only questions in the style ``{\em what}...'', but also ``{\em how}...'', such as ``{\em How} to (always) compute $P$?'', or ``{\em How} to (always) obtain an algorithm for $P$ from an algorithm for $Q$?''. With potential applications in mind, such questions are of course more interesting than their ``{\em what''} style counterparts.

Things are naturally set up so that statements of classical logic turn out to be special cases of computational problems, and classical truth  a special case of computability. Eventually this makes classical logic a conservative fragment of CoL: the language of CoL is a proper extension of that of classical logic, but if we limit the former to the latter, CoL validates nothing more and nothing less than what classical logic does. 

\section{Computability logic versus intuitionistic and linear logics}\label{s2}

 Similarly, intuitionistic and linear logics can also be viewed as fragments of CoL, albeit ``not quite'' conservative ones, as CoL validates certain principles not provable in those logics, even if  there are more similarities than differences. To me this fact indicates that those two logics are incomplete and do not fully correspond to their underlying philosophies and intuitions. 

Let me  take the liberty to philosophize a little bit here. 
I believe the right way to build a new logic is to:
\begin{description}
  \item[(I)] Start with the philosophy and intuitions that we want to capture --- call this {\em informal semantics}.
  \item[(II)]  Then elaborate a {\em formal semantics} that adequately corresponds to the informal semantics.
  \item[(III)] And only after that ask what should be provable and what not in a {\em proof system} for the resulting logic, construct such a system and verify its soundness and completeness.   
\end{description} 
  This is the way classical logic evolved, culminating in G\"{o}del's completeness theorem for first order logic. CoL, too, follows the same pattern. On the other hand, I would say that intuitionistic and linear logics jumped from informal semantics directly into proof systems, skipping the formal semantics phase. 

Take Heyting's intuitionistic logic for instance. Its construction started by looking at  proof systems for classical logic and removing the postulates   that appeared to be wrong from the informal intuitionistic point of view, such as the law of excluded middle. 

Similarly, linear logic was obtained from Gentzen's sequent calculus for classical logic as a result of deleting certain rules obviously incompatible with the resource philosophy of linear logic, such as contraction. 

Yes, in both cases the underlying philosophical and intuitive considerations were sufficient to clearly see that the expelled principles were indeed wrong.   But where is the guarantee that, together with the law of excluded middle or contraction, some innocent, deeply hidden principles did not vanish as well? Idiomatically speaking, where is the guarantee that such a revision of classical logic did not throw out the baby with the bathwater?  And, indeed, I dare to argue that this is exactly what happened. In the case of intuitionistic logic, among such ``babies''  is 

\begin{equation}\label{e1}
\begin{array}{c}(\neg P\rightarrow A\vee B)\wedge(\neg Q\rightarrow  C\vee D)\wedge\neg(P\wedge Q)\rightarrow
 (\neg P\rightarrow A)\vee (\neg P\rightarrow  B) \vee (\neg Q\rightarrow C)\vee(\neg Q\rightarrow D).\end{array} 
\end{equation}
And an example of an innocent victim of rudely rewriting classical logic into linear logic  is  
\begin{equation}\label{e2}
(A\wedge B)\vee(C\wedge D)\rightarrow (A\vee C)\wedge(B\vee D), 
\end{equation}
with its connectives understood in the multiplicative sense. 
I call the latter Blass's principle as Andreas Blass \cite{Bla92} was the first to study it as an example of a game-semantically vaild principle underivable in linear logic.   

Of course some, mostly retroactive, attempts have been made to create formal semantics matching the proof systems of intuitionistic or linear logics. But the reasonable way to go is to match a proof system with a convincing formal semantics rather than vice versa. It is always possible to come up with some formal semantics that matches the target proof system, but the whole question is how adequately and convincingly that semantics captures the philosophy and intuitions underlying the logic.  

When constructing a deductive system, we ask what should be provable in it and what not. An answer to this question stems from the underlying semantics and only semantics, formal or informal: those things should be provable that are semantically valid. Some popular approaches to intuitionistic logic have attempted to explain everything in terms of proofs.  For instance, you can see the meaning of $A\vee B$ explained by saying that this formula should be considered ``good'' (true? provable?) if either $A$ or $B$ can be proven. But the whole point is that we are just trying to understand what should be provable and what not. Trying to justify provability in terms of provability creates a vicious circle.  

Why is taking a shortcut from the earlier described stage (I) directly to stage (III) wrong? Because it is hardly possible to convincingly argue directly that a given proof system corresponds to a given informal semantics. On the other hand, adequacy (soundness and completeness) of a proof system with respect to a formal semantics can be proven mathematically, as both, unlike informal semantics, are mathematical objects. Now you can ask  here: ``OK, but where is then the guarantee that the formal semantics adequately captures the informal semantics and thus the original motivations and philosophy underlying the logic?''. Of course, there is no guarantee, as this cannot be proven mathematically. But it is easier to argue that the two match each other (when they really do) because both are semantics.  Comparing apples with apples is easier than comparing them with oranges.  

I have been  pushing forward the above points since long ago. While having heard the angry ``How dare you!" many times from sympathizers of intuitionistic or linear logics, I am still waiting to see some more convincing attempts  to refute them.  

Summarizing much of what has been said in this section, my favorite excerpt from \cite{Japfin}, not without sarcasm, notes:

\begin{quote}{\em The reason for the failure of the principle of excluded middle in CoL is not that this principle ... is not included in its axioms. Rather, the failure of this principle is exactly the reason why it, or anything entailing it, would not be among the axioms of a sound system for CoL.}
\end{quote} 

\section{Computational problems as games}\label{s3}

Anyway, what is computability? Before trying to answer or even ask this question, one should first understand what a computational problem is, for computability is a property of computational problems. So, what is a {\em computational problem}? According to Church,  a computational problem is nothing but a {\em function} (to be computed). That is, the task of systematically generating the values of that function at different arguments. The tradition of seeing computational problems as functions has since  firmly established in theoretical computer science. Such an approach, however, as acknowledged by Turing  \cite{Tur36} himself,  is too narrow.  Most tasks performed by computers are {\em interactive}, far from being as simple as just receiving  an input and generating an output. 
For instance, take a look at the work of a network server. It is in fact an infinite process, with signals moving back and forth between it and its environment in a not quite synchronized or regulated fashion, affecting not only current events but some future events as well. Such tasks are not always reducible to functions, at least reducible in some ``nice'' way. We need something more here, a more general concept  to be able to adequately  model complex tasks performed by computers.

Such ``something more'' for us are {\em games}: a computational problem is a game between a machine, denoted $\top$, and its environment, denoted $\bot$. Then {\em computability} is understood as existence of a machine which always wins the game, i.e., wins it no matter how the environment acts. 
In this presentation I am not giving you any formal definitions, including  definitions of our concepts of games or game-playing. But such definitions, of course, do exist. 

Even though often it is us who act in the role of $\bot$, we are fans of $\top$ rather than $\bot$. That is because $\top$ (machine) is a tool, and its losing the game would mean failing to perform the task it was supposed to perform for us.  The behavior (game-playing strategy) of $\top$, as the word ``machine'' suggests, should be algorithmic as it is a mechanical device. On the other hand, there are no restrictions on the behavior of $\bot$, as the latter represents a capricious user, the blind forces of nature or the devil himself (and you can't ask the devil to only follow algorithmic strategies). 

Games can be visualized as trees in the style of \mbox{Figure 1.} Vertices of such a tree represent positions in the game, and edges --- their labels, that is --- represent legal moves, prefixed with $\top$ or $\bot$ to indicates which player can make the move. On the other hand, the label $\top$ or $\bot$ of a vertex indicates which  player is considered to be the winner if the game ends in the corresponding position. The game can end anywhere, it does not have to continue to the ``end'': after all, some branches can be infinite and thus there will be nothing that could be understood as the ``end''. So, if the machine made the move $\alpha$ in the  game of \mbox{Figure 1,} the environment responded with $\gamma$ and no further moves were made, the machine loses as the corresponding vertex of the tree is 
$\bot$-labeled.
  
\begin{center}
\begin{picture}(322,210)

\put(159,192){\circle{16}}
\put(155,188){$\bot$}
\put(159,184){\line(-3,-1){104}}
\put(80,163){{\tiny $\top$}{\footnotesize $\alpha$}}
\put(159,184){\line(0,-1){34}}
\put(161,163){{\tiny $\bot$}{\footnotesize $\beta$}}
\put(159,184){\line(3,-1){104}}
\put(228,163){{\tiny $\bot$}{\footnotesize $\gamma$}}

\put(52,142){\circle{16}}
\put(48,138){$\top$}
\put(52,134){\line(-3,-2){51}}
\put(8,113){{\tiny $\bot$}{\footnotesize $\beta$}}
\put(52,134){\line(3,-2){51}}
\put(87,113){{\tiny $\bot$}{\footnotesize $\gamma$}}
\put(159,142){\circle{16}}
\put(155,138){$\top$}
\put(159,134){\line(0,-1){34}}
\put(161,113){{\tiny $\top$}{\footnotesize $\alpha$}}
\put(266,142){\circle{16}}
\put(262,138){$\bot$}
\put(266,134){\line(-3,-2){51}}
\put(221,113){{\tiny $\top$}{\footnotesize $\alpha$}}
\put(266,134){\line(0,-1){34}}
\put(268,113){{\tiny $\top$}{\footnotesize $\beta$}}
\put(266,134){\line(3,-2){51}}
\put(299,113){{\tiny $\top$}{\footnotesize $\gamma$}}

\put(0,92){\circle{16}}
\put(-4,88){$\top$}
\put(104,92){\circle{16}}
\put(100,88){$\bot$}
\put(104,84){\line(-1,-3){11}}
\put(85,63){{\tiny $\top$}{\footnotesize $\beta$}}
\put(104,84){\line(1,-3){11}}
\put(112,63){{\tiny $\top$}{\footnotesize $\gamma$}}
\put(159,92){\circle{16}}
\put(155,88){$\top$}
\put(214,92){\circle{16}}
\put(210,88){$\bot$}
\put(214,84){\line(-1,-3){11}}
\put(195,63){{\tiny $\top$}{\footnotesize $\beta$}}
\put(222,63){{\tiny $\top$}{\footnotesize $\gamma$}}
\put(214,84){\line(1,-3){11}}
\put(266,92){\circle{16}}
\put(262,88){$\top$}
\put(266,84){\line(0,-1){34}}
\put(268,63){{\tiny $\top$}{\footnotesize $\alpha$}}
\put(318,92){\circle{16}}
\put(314,88){$\bot$}
\put(318,84){\line(0,-1){34}}
\put(320,63){{\tiny $\top$}{\footnotesize $\alpha$}}

\put(92,42){\circle{16}}
\put(88,38){$\top$}
\put(116,42){\circle{16}}
\put(112,38){$\bot$}
\put(202,42){\circle{16}}
\put(198,38){$\top$}
\put(226,42){\circle{16}}
\put(222,38){$\bot$}
\put(266,42){\circle{16}}
\put(262,38){$\top$}
\put(318,42){\circle{16}}
\put(314,38){$\bot$}

\put(92,10){{\bf Figure 1:} A  game of depth 3}
\end{picture}
\end{center}

Games in logic have been studied by many authors, but our understanding of games is apparently unique in that it does not impose any regulations on the order in which the players should or could move, and permits positions where both players have legal moves. For instance, the root position of the game of Figure 1, as we see, allows either player to move. In that position, a move (if any) will be made  by the player which can or want to act faster. 

It turns out that, in the sort of games we consider, the relative speed of either player does  not matter. Namely, it never hurts a player to postpone making moves and let the adversary go first whenever possible. Such games are said to be {\em static}, and they are defined by imposing a certain technical yet simple condition on games. Striving to keep this presentation non-technical, I will not discuss that condition here. Suffice it to say that all  ``pure'' (speed-independent) interactive problems turn out to be static, and the class of static games is closed under all game operations studied in CoL. The game of \mbox{Figure 1} is static, in which the machine has a winning strategy. An interactive algorithm that guarantees the machine a win reads as follows: 
\begin{quote} {\em Regardless of what the adversary is doing or has done, go ahead and make move $\alpha$; make $\beta$ as your second (and last) move if and when you see that the adversary has made move $\gamma$, no matter whether this happened before or after your first move. } 
\end{quote}
It is left as an exercise for the reader to see that $\top$, following this interactive algorithm (strategy), wins no matter what and how fast $\bot$ does.

Computational problems in the traditional sense, i.e. functions, are static games of depth 2 of the kind seen in \mbox{Figure 2}.

\begin{center}
\begin{picture}(322,160)

\put(175,142){\circle{16}}
\put(-4,109){\scriptsize Input}
\put(171,138){$\top$}
\put(175,134){\line(-3,-1){106}}
\put(83,109){\tiny $\bot 0$}
\put(175,134){\line(0,-1){34}}
\put(163,109){\tiny $\bot 1$}
\put(175,134){\line(3,-1){106}}
\put(226,109){\tiny $\bot 2$}
\put(175,134){\line(6,-1){119}}
\put(285,109){\Huge ...}

\put(-4,59){\scriptsize Output}
\put(65,92){\circle{16}}
\put(61,88){$\bot$}
\put(30,59){\tiny $\top 0$}
\put(65,84){\line(-1,-3){11}}
\put(46,59){\tiny $\top 1$}
\put(65,84){\line(-1,-1){34}}
\put(62,59){\tiny $\top 2$}
\put(65,84){\line(1,-3){11}}
\put(77,59){\tiny $\top 3$}
\put(65,84){\line(1,-1){34}}
\put(95,59){\large ...}
\put(65,84){\line(3,-2){34}}

\put(29,42){\circle{16}}
\put(25,38){$\bot$}
\put(53,42){\circle{16}}
\put(49,38){$\top$}
\put(77,42){\circle{16}}
\put(73,38){$\bot$}
\put(101,42){\circle{16}}
\put(97,38){$\bot$}

\put(175,92){\circle{16}}
\put(171,88){$\bot$}
\put(140,59){\tiny $\top 0$}
\put(175,84){\line(-1,-3){11}}
\put(156,59){\tiny $\top 1$}
\put(175,84){\line(-1,-1){34}}
\put(172,59){\tiny $\top 2$}
\put(175,84){\line(1,-3){11}}
\put(187,59){\tiny $\top 3$}
\put(175,84){\line(1,-1){34}}
\put(205,59){\large ...}
\put(175,84){\line(3,-2){34}}

\put(139,42){\circle{16}}
\put(135,38){$\bot$}
\put(163,42){\circle{16}}
\put(159,38){$\bot$}
\put(187,42){\circle{16}}
\put(183,38){$\top$}
\put(211,42){\circle{16}}
\put(207,38){$\bot$}

\put(285,92){\circle{16}}
\put(281,88){$\bot$}
\put(250,59){\tiny $\top 0$}
\put(285,84){\line(-1,-3){11}}
\put(266,59){\tiny $\top 1$}
\put(285,84){\line(-1,-1){34}}
\put(282,59){\tiny $\top 2$}
\put(285,84){\line(1,-3){11}}
\put(297,59){\tiny $\top 3$}
\put(285,84){\line(1,-1){34}}
\put(315,59){\large ...}
\put(285,84){\line(3,-2){34}}

\put(249,42){\circle{16}}
\put(245,38){$\bot$}
\put(273,42){\circle{16}}
\put(269,38){$\bot$}
\put(297,42){\circle{16}}
\put(293,38){$\bot$}
\put(321,42){\circle{16}}
\put(317,38){$\top$}

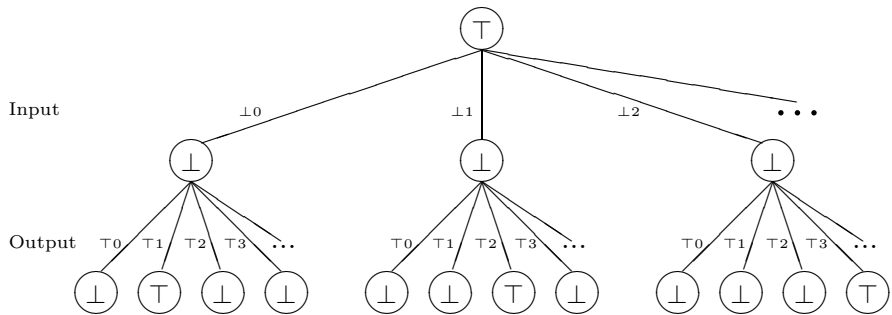
\put(55,10){{\bf Figure 2:} The successor function as a game}
\end{picture}
\end{center}

In such a game, the upper level edges represent possible inputs provided by the environment, so they are $\bot$-labeled. The lower level edges represent possible outputs generated by the machine, so they are $\top$-labeled. The root is $\top$-labeled because it corresponds to the situation where nothing happened, namely, no input was provided by the environment. The machine has nothing to answer for in this case, so it wins. The middle level nodes are $\bot$-labeled because they correspond to situations where there was an input but the machine failed to generate an output, so the machine loses. Each group of the bottom level nodes has exactly one $\top$-labeled node, because a function has exactly one (correct) value at each argument. It is not hard to see that the  particular game of \mbox{Figure 2} represents the successor function $x+1$: if the input is $0$, the machine, in order to win, should generate the output $1$, if the input is $1$, the output should be $2$, etc. 

Now CoL rhetorically asks  why limit ourselves only to trees of the kind seen in \mbox{Figure 2}.   
First of all, we may want to allow branches to be longer than $2$, or even infinite to be able to model long or infinite tasks performed by computing machines.  And why not allow all sorts of distributions of $\top$ and $\bot$ in nodes or on edges? For instance, consider the task of computing the function $3/x$. It would be natural to make the node to which the input $0$ takes us not $\bot$-labeled, but $\top$-labeled. Because the function is not defined at $0$, so the machine cannot be held responsible for failing  to generate an output on such an input.  

It makes sense to generalize computational problems not only in the direction of increasing their depths, but also decreasing. Games of depth $0$ are said to be {\em elementary}. These are games with no legal moves (the game ``tree'' is just its root), and thus games where one of the players automatically wins by doing nothing. We understand true atomic sentences of classical logic such as $2\times 2=4$ or $\top$ as the elementary game automatically won by the machine, and false sentences such as $2\times 2=5$ or $\bot$ as the elementary game lost by the machine. Note the two different yet related meanings of the symbols $\top$ and $\bot$ in CoL: depending on the context, such a symbol stands either for the corresponding elementary game, or the player which wins that game. 

Thus, classical propositions for us are nothing but elementary games. This generalizes to predicates in the standard way. In classical logic, predicates can be thought of as ``propositions that (may) depend on variables''. Similarly, we allow ``games that (may) depend on variables'', with predicates being nothing but elementary sorts of such games. As a result, classical logic becomes a special case of CoL --- CoL where only elementary games are allowed. 

\section{Choice operators}
Logical operators in CoL stand for operations on games. There is a whole zoo of them, with (at least) four sorts of conjunction and disjunction as well as universal and existential quantifiers,  a bunch of so called recurrence (repetition) operations and corresponding implication-style and negation-style operations, and more. In this short presentation we shall only look at the following subset of the logical operators studied in CoL: 
\[\neg,\wedge,\vee,\rightarrow,\cla,\cle,\adc,\add,\ada,\ade,\st,\cost,\intimpl,\brneg.\] 
Using the classical notation for the  first six of these is no accident. They are conservative generalizations of their classical counterparts from elementary games to all games. Conservative in the sense that, when applied to elementary games (propositions, predicates) only, their extensional meanings and logical behavior turn out to be exactly classical. This is how classical logic naturally becomes a special (elementary)  fragment of CoL. 

We start with the  {\em choice connectives} $\adc$ (conjunction) and $\adc$ (disjunction). The way they combine two games $A$ and $B$ to get the new game $A\adc B$ or $A\add B$ is depicted in \mbox{Figure 3}.

  \begin{center}
\begin{picture}(226,123)

\put(26,105){$A\adc B$}
\put(41,87){\circle{16}}
\put(37,83){$\top$}
\put(41,79){\line(-1,-1){34}}
\put(7,58){\scriptsize $\bot 0$}
\put(41,79){\line(1,-1){34}}
\put(64,58){\scriptsize $\bot 1$}

\put(0,34){$A$}
\put(70,34){$B$}

\put(176,105){$A\add B$}
\put(191,87){\circle{16}}
\put(187,83){$\bot$}
\put(191,79){\line(-1,-1){34}}
\put(157,58){\scriptsize $\top 0$}
\put(191,79){\line(1,-1){34}}
\put(214,58){\scriptsize $\top 1$}

\put(150,34){$A$}
\put(220,34){$B$}

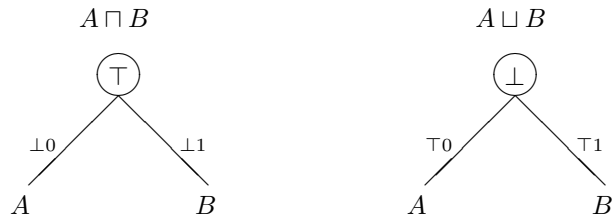
\put(10,10){{\bf Figure 3:} Choice conjunction and disjunction}
\end{picture}
\end{center} 

As we see, $A\adc B$ is the game where the first legal move is (only) by the environment. Such a move should be either $0$ or $1$. If move $0$ is made, the game ``turns into'' $A$, in the sense that it continues --- and the winner is determined --- according to the rules of $A$. Similarly for $B$ in the case of move $1$. Intuitively, making  move $0$ or $1$ means choosing between the left disjunct and the right disjunct. Making such a choice is not only a privilege of the environment, but also an obligation: as seen in the picture, the root of $A\adc B$ is $\top$-labeled, meaning that the environment loses if it fails to make an initial move/choice.  

$A\add B$ is fully symmetric/dual to $A\adc B$: in it, it is the machine rather than the environment who makes the initial choice and who loses if no choice is made. 

For  simplicity, let us agree that  the universe of discourse is always $\{0,1,2,\cdots\}$. If so, the {\em choice universal quantification} $\ada xA(x)$ (note that $\ada$ is larger than $\adc$) can be understood as the infinite choice conjunction $A(0)\adc A(1)\adc A(2)\adc\cdots$, and the {\em choice existential quantification} $\ade xA(x)$ as the infinite disjunction $A(0)\add A(1)\add A(2)\add\cdots$. So, now a choice is made not just between $0$ or $1$, but among $0,1,2,\cdots$, as shown in \mbox{Figure 4}.  
   
 \begin{center}
\begin{picture}(226,125)

\put(28,107){$\ada xA(x)$}
\put(46,89){\circle{16}}
\put(42,85){$\top$}
\put(46,81){\line(-1,-1){34}}
\put(11,58){\scriptsize $\bot 0$}
\put(46,81){\line(1,-1){34}}
\put(54,58){\scriptsize $\bot 2$}
\put(46,81){\line(0,-1){33}}
\put(34,58){\scriptsize $\bot 1$}
\put(46,81){\line(3,-2){30}}
\put(76,58){\bf \ldots}

\put(0,34){$A(0)$}
\put(35,34){$A(1)$}
\put(70,34){$A(2)$}

\put(178,107){$\ade xA(x)$}
\put(196,89){\circle{16}}
\put(192,85){$\bot$}
\put(196,81){\line(-1,-1){34}}
\put(161,58){\scriptsize $\top 0$}
\put(196,81){\line(1,-1){34}}
\put(204,58){\scriptsize $\top 2$}
\put(196,81){\line(0,-1){33}}
\put(184,58){\scriptsize $\top 1$}
\put(196,81){\line(3,-2){30}}
\put(226,58){\bf \ldots}

\put(150,34){$A(0)$}
\put(185,34){$A(1)$}
\put(220,34){$A(2)$}

\put(52,10){{\bf Figure 4:} Choice quantifiers}
\end{picture}
\end{center}

Having these operators in the language, we may now conveniently express standard computational problems (and beyond) without drawing trees.  So, for instance, the problem of computing the successor function depicted in \mbox{Figure 2} can be simply written as $\ada x\ade y(y=x+1)$. In this game, the first move --- for instance $2$ --- is by the environment. Intuitively, this can be seen as asking the machine the question ``What is the successor of $2$?''. The game continues as $\ade y(y=2+1)$. The next move --- say $3$ --- is by the machine, which amounts to saying that $3$ is the successor of $2$. The game is now brought down (``continues as'') $3=2+1$. This is an elementary game with no further moves, and the machine has won because $3=2+1$ is true. Had the machine made the move $4$ instead of $3$, or no move at all, it would have lost.  

Rather similarly, the problem of deciding a predicate $p$ is expressed by $\ada x\bigl(\neg p(x)\add p(x)\bigr)$. 

\section{Negation}
{\em Negation}  $\neg$ is an operation which flips the roles of the two players, turning $\top$'s wins and legal moves into $\bot$'s wins and legal moves, and vice versa. For instance, if $\chess$ is the game of chess 
from the point of view of the white player, then $\neg\chess$ is the same game as seen by the black player. Figure 5 illustrates how applying $\neg$ to a game $A$ generates the exact ``negative image'' of $A$, with $\top$ and $\bot$ interchanged both in the nodes and on the arcs of the game tree.
 
\begin{center}
\begin{picture}(311,178)

\put(49,160){$A$}
\put(53,142){\circle{16}}
\put(49,138){$\top$}
\put(53,134){\line(-1,-1){34}}
\put(20,113){\scriptsize $\bot 0$}
\put(53,134){\line(1,-1){34}}
\put(76,113){\scriptsize $\bot 1$}

\put(17,92){\circle{16}}
\put(13,88){$\bot$}
\put(-2,62){\scriptsize $\top 0$}
\put(17,84){\line(-1,-3){11}}
\put(17,84){\line(1,-3){11}}
\put(25,62){\scriptsize $\top 1$}
\put(4,42){\circle{16}}
\put(0,38){$\top$}
\put(28,42){\circle{16}}
\put(24,38){$\bot$}

\put(88,92){\circle{16}}
\put(84,88){$\bot$}
\put(69,62){\scriptsize $\top 0$}
\put(88,84){\line(-1,-3){11}}
\put(88,84){\line(1,-3){11}}
\put(96,62){\scriptsize $\top 1$}
\put(75,42){\circle{16}}
\put(71,38){$\bot$}
\put(99,42){\circle{16}}
\put(95,38){$\top$}

\put(244,160){$\neg A$}
\put(253,142){\circle{16}}
\put(249,138){$\bot$}
\put(253,134){\line(-1,-1){34}}
\put(220,113){\scriptsize $\top 0$}
\put(253,134){\line(1,-1){34}}
\put(276,113){\scriptsize $\top 1$}

\put(217,92){\circle{16}}
\put(213,88){$\top$}
\put(198,62){\scriptsize $\bot 0$}
\put(217,84){\line(-1,-3){11}}
\put(217,84){\line(1,-3){11}}
\put(225,62){\scriptsize $\bot 1$}
\put(204,42){\circle{16}}
\put(200,38){$\bot$}
\put(228,42){\circle{16}}
\put(224,38){$\top$}

\put(288,92){\circle{16}}
\put(284,88){$\top$}
\put(269,62){\scriptsize $\bot 0$}
\put(288,84){\line(-1,-3){11}}
\put(288,84){\line(1,-3){11}}
\put(296,62){\scriptsize $\bot 1$}
\put(275,42){\circle{16}}
\put(271,38){$\top$}
\put(299,42){\circle{16}}
\put(295,38){$\bot$}

\put(99,10){{\bf Figure 5:} Negation}
\end{picture}
\end{center}

Obviously if $A$ is a true proposition, i.e., an elementary game automatically won by the machine, then $\neg A$ remains an elementary game but now lost by the machine; in other words, $\neg A$ is a false proposition. This is exactly what was meant when promising that the meaning of $\neg$, or any other operator for which we use classical notation,  is exactly classical when limited to elementary games. 

It can be easily seen that the games $\neg\neg A$ and $A$ are identical: switching the roles twice brings each player to its original status. Similarly, it can be seen that $\neg$ interacts with choice operations in the kind old DeMorgan fashion. E.g., $\neg(A\adc B)=\neg A\add\neg B$. Looking back at \mbox{Figure 5}, notice that the game $A$ shown there is nothing but $(\top\add\bot)\adc(\bot\add\top)$, and $\neg A$ is its DeMorgan dual $(\bot\adc\top)\add(\top\adc\bot)$.

\section{Parallel connectives}  

The operations $\wedge$ and $\vee$ are called {\em parallel conjunction} and {\em parallel disjunction}. Unlike their choice counterparts $A\adc B$ and $A\add B$, in $A\wedge B$ or $A\vee B$ no choice between $A$ and $B$ is made by either player. Rather, the play proceeds in parallel in both components. To win in $A\wedge B$, the machine should win in both $A$ and $B$, while for winning $A\vee B$ winning in just one of the two components is sufficient. 

Consider, for instance,  $\chess \wedge \chess$. This is in fact a play on two boards, where $\top$ plays white on both boards. Perhaps it plays against two adversaries: Peter and Paul, though, for $\top$, they together form just what it calls the (one) environment. In order to win, $\top$ needs to defeat Peter on the left board {\em and} Paul on the right board. The first move in this compound game is definitely by $\top$, as the opening move is by the white player on both boards. But, after $\top$ makes its first move, say against Peter, the situation changes. Now both $\top$ and its environment naturally have legal moves. Namely, $\top$ has a legal move against Paul, while Peter (and thus the environment from $\top$'s point of view) also has a legal move in response to $\top$'s initial move. It would be unnatural here to impose some regulations regarding which player can go next. This is why CoL's understanding of games does not insist that in each position only either $\top$ or $\bot$ (but not both) should be allowed to move. 

To appreciate the difference between the choice and the parallel sorts of connectives, let us compare the two games $\neg \chess\add \chess$ and $\neg \chess \vee \chess$. We assume that draw outcomes are ruled out in $\chess$, and the player who fails to make a move on his turn is considered to have lost. Imagine I am playing in the role of $\top$, and the world champion Kasparov in the role of $\bot$. In $\neg \chess\add \chess$, I have a choice between playing on the left board ($\neg \chess$) or on the right board ($\chess$). That is, I get to decide whether I want to play black or play white.  After  such a choice is made, I have to defeat Kasparov on the chosen board, while the other board is discarded. Obviously I stand no chance to win, regardless of whether I choose to play black or white.  On the other hand, I can easily beat Kasparov in $\neg \chess \vee \chess$. This is a parallel play on two boards. At the beginning, both Kasparov and I have legal moves: Kasparov on the left board where he is playing white, and I on the right board. Rather than hurrying to make an opening move, I wait to let Kasparov move first. If he, too, chooses to do nothing, then I win due to being the winner on the left board. Now suppose Kasparov makes his opening move on the left board. Can you guess how I should respond? Yes, by making the exact same move on the right board. I wait again. Whatever move Kasparov makes on the right board in response, I copy that move back on the left board. And so on. By using this copy-cat strategy, I am in fact letting Kasparov play against himself. Eventually, both he and I are guaranteed to win on one board and lose on the other. Since this is a disjunction, having won in one of the disjuncts makes me the winner in the overall game.  

In general, the law of excluded middle \mbox{``$\neg A$ OR $A$''} is invalid in CoL with OR understood as $\add$ but valid when OR is understood as $\vee$: one can prove that, while the above seen copy-cat strategy wins all games of the form $\neg A\vee A$, for some $A$  no machine can win $\neg A\add A$ against a sufficiently smart adversary. 

\section{Putting things where they belong}
What is meant by ``Giving Caesar what belongs to Caesar'' (... and God what belongs to God) in the title of  this article? The twentieth century has witnessed endless and fruitless fights between the classically-minded and the constructivistically-minded regarding whether the law of excluded middle should be accepted or rejected. It is obvious that the two schools of thought were talking about two very different meanings of disjunction. Yet,   for some strange reason, they chose the same symbol $\vee$ for both, and then started arguing with each other. Not quite serious I would say. CoL neutralizes this and similar controversies by putting things where they belong. And, as pointed out in Section \ref{s2}, it does so {\em semantically}, not because it allows or forbids them among the postulates of some purportedly ``right'' deductive system.  

Give the classically minded what belongs to the classically-minded ($\vee$), and the constructivists what belongs to the constructivists ($\add$)! 

\begin{itemize}
  \item Yes, classical logic is right: $\neg A\vee A$ is indeed valid. 
  \item Yes, intuitionistic logic is right: $\neg A\add A$ is indeed invalid.
  \end{itemize} 
No subject for arguing! 

The classical tautology $(\neg A\wedge \neg A)\vee A$ fails in CoL unless $A$ is stipulated to be elementary.    
Observe that, at least, the copy-cat trick  used earlier in our winning strategy for  $\neg \chess\vee \chess$ no longer works for the ``similar'' $(\neg \chess\wedge \neg \chess)\vee \chess$. I can try to copy Kasparov's moves in $\chess$ within both conjuncts of $\neg \chess \wedge \neg \chess$ and vice versa. However, Kasparov may start acting in different ways in these two conjuncts, and then, at best, I will be able to synchronize only one of them with $\chess$. It is then possible that eventually I lose in $\chess$ and in the unsynchronized conjunct of    $\neg \chess \wedge \neg \chess$, which makes me lose in the overall game $(\neg \chess\wedge \neg \chess)\vee \chess$. Anyway, classical logic accepts the  principle $(\neg A\wedge \neg A)\vee A$ and linear logic rejects it (with $\wedge,\vee$ seen as multiplicatives). Which one is ``right''? 

The formal language of pure CoL has two sorts of nonlogical atoms: {\em elementary} and {\em general}. Elementary atoms are meant to be interpreted as elementary games, and general atoms as any games, elementary or not. We use the lowercase $p,q,\cdots$ for elementary atoms and the uppercase $P,Q,\cdots$ for general atoms.   

And, again, Caesar is being given what belongs to Caesar and God what belongs to God. The semantics of CoL classifies:

\begin{itemize}
  \item $(\neg p\wedge\neg p)\vee p$ as valid. Yes, classical logic is right!
  \item  $(\neg P\wedge\neg P)\vee P$ as invalid. Yes, linear logic is right!
  \end{itemize}
(As for the earlier discussed law of excluded middle, both $\neg P\vee P$ and $\neg p\vee p$ are valid and both  $\neg P\add P$ and $\neg p\add p$ are invalid.)

From CoL's perspective, classical logic differs from intuitionistic logic in its understanding of logical constants (operators), and differs from linear logic in its understanding of logical variables (nonlogical atoms). 

\section{Reduction}
The {\em implication} operation $\rightarrow$ is defined in the standard way by \[A\rightarrow B\  =_{def} \neg A\vee B.\] The intuition associated with this operation is that of a {\em reduction} of the consequent to the antecedent. Since $A$ is negated here and thus the roles of the two players are interchanged in it, $A$ can be seen by $\top$ as an environment-provided {\em resource} rather than a task. Namely, $\top$ can observe how the environment is playing in $A$ and use that information in its play in $B$. The task of $\top$ is to win $B$ as long as the environment wins $A$; in other words, to solve problem $B$ as long as the environment is (correctly) solving problem $A$. 

To get a feel of $\rightarrow$ as a reduction operation, consider the game 
\[\begin{array}{c}\ada x\ade y \bigl(y=\father(x)\bigr)\wedge \ada x\ade y \bigl(y=\mother(x)\bigr)
\rightarrow \ \ada x\ade y \bigl(y=\nainai (x)\bigr),\end{array}\]
where $\father(x)$ is the function  ``$x$'s father'', and similarly for $\mother(x)$ and $\nainai(x)$. Here, the task $\top$ is facing is telling the name of an arbitrary person's paternal grandmother while the environment (correctly) tells the name of an arbitrary person's father and the name of an arbitrary person's mother. In other words, this is the problem of reducing the paternal grandmotherhood problem to the fatherhood and motherhood problems. Winning this game is easy and does not require any knowledge of anyone's relative's names.   Here is a strategy for $\top$: Wait till $\bot$ makes a move $a$ in the consequent (if not, $\top$ wins automatically). Intuitively, such a move amounts to asking $\top$ the question ``Who is $a$'s paternal grandmother?''. Make the same move $a$ in the first conjunct of the antecedent, i.e., ask  the counterquestion ``Who is $a$'s father?''. $\bot$ will have to answer correctly, or else it loses. Let us say $\bot$'s answer/move is $b$. Make the same move $b$ in the second conjunct of the antecedent, thus asking $\bot$ to tell who $b$'s mother is. 
$\bot$, again, will have to provide the correct answer, let us say $c$. Now, by making the same move $c$ in the consequent, i.e.,  answering ``$c$'' to $\bot$'s original question regarding $a$'s paternal grandmother, $\top$ wins: $c$ is indeed $a$'s paternal grandmother  (unless the environment lied in the antecedent about $a$'s father or $b$'s mother, but in that case, as already noted, $\top$ is no longer responsible for anything). 

\section{Blind quantifiers} 
The operations $\cla$ and $\cle$, called {\em blind quantifiers}, conservatively generalize their classical counterparts, just like $\neg,\wedge,\vee,\rightarrow$ do. Unlike the choice quantifiers, there are no moves associated with $\cla$ or its dual $\cle$. Playing $\cla xA(x)$ or $\cle xA(x)$ means playing $A(x)$ ``blindly'', without knowing the value of $x$ as the latter is not specified by either player. In order to win $\cla xA(x)$ (resp. $\cle xA(x)$), $\top$ needs to play $A(x)$ in such a way that it wins for all (resp. at least one) possible values of $x$. 

An alternative intuitive characterization of $\cla xA(x)$ and $\cle xA(x)$ would be that, in these games, a third party chooses a value for $x$ but never shows it to either player. In order to win $\cla xA(x)$ (resp. $\cle xA(x)$),  $\top$  (resp. $\bot$) needs to play $A(x)$ in a way that guarantees success regardless of what that chosen value might have been. 

Let us compare the games \[\ada x\bigl(\even(x)\add \odd(x)\bigr) \mbox{ \ and \ } \cla x\bigl(\even(x)\add \odd(x)\bigr).\]
$\ada x\bigl(\even(x)\add \odd(x)\bigr)$, which is a game of depth 2, is easy to win: wait till the adversary selects a value $m$ for $x$; if $m$ is even, respond  by choosing the left disjunct of $\even(m)\add \odd(m)$, otherwise respond by choosing the right disjunct, and rest your case.  On the other hand, $\cla x\bigl(\even(x)\add \odd(x)\bigr)$ is  a game of depth 1, and it is impossible to win. Here the value of $x$ is not specified by the adversary or whoever for that matter,  yet you should do the impossible task of choosing between $\even(x)$ and $\odd(x)$  so that all of the elementary games/propositions $\even(0), \even(1), \even(2),\cdots $ (if you chose $\even(x)$) or $\odd(0), \odd(1), \odd(2),\cdots $ (if you chose $\odd(x)$) are won/true. 

This should not suggest than all $\cla$-games are unwinnable. Consider 
\[\cla x\Bigl(\even(x)\add \odd(x)\rightarrow \ada y\bigl( \even(x+y)\add \odd(x+y)\bigr)\Bigr).\]  
Here, given a number chosen by the environment for $y$, let us say $5$, in order to tell whether $x+5$ is even or odd it is not necessary to know the actual value of $x$. Rather, just knowing whether $x$ is even or odd is sufficient. And, luckily, this piece of information on $x$ will have to be provided by the environment as mandated by $\even(x)\add \odd(x)$ in the antecedent. If the environment claims that $x$ is even, then $\top$ chooses 
$\odd(x+5)$ and wins; otherwise, it chooses $\even(x+5)$.

$\cla$ can be seen to be stronger than $\ada$,  in the sense that the semantics of CoL validates the principle $\cla xA(x)\rightarrow \ada xA(x)$ but not its contrapositive. This means that $\ada xA(x)$ is reducible to $\cla xA(x)$ but not vice versa. Symmetrically, $\cle$ is weaker than $\ade$. 

Speaking philosophically, choice quantifiers are {\em constructive} versions of their blind counterparts. 
While not as popular as the law of excluded middle, $\cle x\cla y\bigl(p(x)\vee\neg p(y)\bigr)$ is another example of a valid principle of classical logic which, however, is not valid in any constructive sense, and not provable in intuitionistic logic. Again giving Caesar what belongs to Caesar, CoL unsurprisingly establishes:
\begin{itemize}
  \item Both $\cle x\cla y\bigl(p(x)\vee\neg p(y)\bigr)$ and $\cle x\cla y\bigl(P(x)\vee\neg P(y)\bigr)$ are valid. Yes, classical logic is right!
  \item Both $\ade x\ada y\bigl(p(x)\vee\neg p(y)\bigr)$ and $\ade x\ada y\bigl(P(x)\vee\neg P(y)\bigr)$ are invalid. Yes, intuitionistic logic is right!
  \end{itemize}
On the other hand, the valid principle $\cla y\cle x\bigl(p(x)\vee\neg p(y)\bigr)$ of classical logic is commonly recognized to be valid in every reasonable constructive sense, and is provable in intuitionistic logic. As expected, CoL validates this principle with both (blind and choice) sorts of quantifiers and both (elementary and general) sorts of atoms.

\section{Recurrences}

Out of several types of so called {\em recurrence} operations studied within the framework of CoL, here we shall only take a look at  {\em branching recurrence} $\st$. Its dual {\em corecurrence} operation $\cost$ can simply be understood as $\neg\st\neg$.  When applied to a game $G$, $\st$ turns it into a game playing which means repeatedly playing $G$. When $G$ is seen as a resource (e.g., when it is in the antecedent of an implication), $\st$ generates multiple ``copies''  of $G$, thus making $G$ a reusable/recyclable resource. 

In classical logic, this sort of an operation  would be meaningless, because classical logic  is resource-blind, seeing  no difference between one and many copies of $G$. In the resource-conscious CoL, however, recurrence operations are not only meaningful, but also necessary to achieve a satisfactory level of expressiveness and realize CoL's potential and ambitions. Hardly any computer program is used only once; rather, it is run over and over again. Loops within such programs also assume multiple repetitions of the same subroutine. In general, the tasks performed in  real life by computers, robots or humans are typically recurring ones or involve recurring subtasks.

Let me use our old friend $\chess$ to explain the meaning of $\st$. A play of $\st \chess$ starts as an ordinary play of $\chess$. At any time, however, the environment may decide to split the current position into two identical ones, thus creating two runs of $\chess$ out of one that have a common past  but possibly diverging futures. From that point on, the play continues on two boards. At any time, the environment can again create two identical copies of the then-current position on either board, and the play correspondingly continues on three boards. The environment can keep splitting  positions in this fashion, creating more and more sessions of $\chess$ to be played in parallel. Eventually, $\top$ will be considered the winner if it wins in all of those sessions.  $\cost \chess$ is similar, with the difference that now splitting positions is $\top$'s privilege, and $\top$ wins if it wins in at least one of the multiple sessions of $\chess$. 

\section{Brimplication} 

The implication-style operation $\intimpl$, called {\em brimplication} (``br'' for ``branching''), is  defined by \[A\intimpl B\  =_{def} \st A\rightarrow B.\] Intuitively $A\intimpl B$, just like $A\rightarrow B$, is a problem of reducing $B$ to $A$. The difference between the two reduction operations is that, while in  $A\rightarrow B$ the machine has a single  copy of $A$ available as an environment-provided informational resource for solving $B$, in $A\intimpl B$ the resource $A$ --- as well as any game/position it has evolved to --- can be duplicated and reused any number of times. As a result, $A\intimpl B$ is easier for $\top$ to win than $A\rightarrow B$ because,  as a resource, the antecedent of $A\intimpl B$ is stronger (very much so) than the antecedent of $A\rightarrow B$.  While being the most {\em basic} sort of reduction allowing us to naturally define $\intimpl$ or other reduction-style operations, $\rightarrow$ is a stricter and thus less general operation of reduction than $\intimpl$.  In fact, according to Thesis \ref{thesis} below,  $\intimpl$ is {\em the most general} sort of reduction. 

We say that a problem/game $B$ is {\em brimplicatively reducible} to a problem $A$ iff there is a machine with a winning strategy for  $A\intimpl B$. 
    
\begin{thesis}\label{thesis} {\em Brimplicative reducibility is an adequate mathematical counterpart of our intuition of reducibility in the weakest --- and hence the most general --- algorithmic sense possible.  Namely, for all games/problems $A$ and $B$, we have:
\begin{description}
  \item[(I):] Whenever  $B$ is brimplicatively reducible  $A$, $B$ is also algorithmically reducible to $A$ according to everyone's reasonable intuition.
  \item[(II):] Whenever   $B$ is algorithmically reducible to  $A$ according to everyone's reasonable intuition, $B$ is also brimplicatively reducible to $A$. 
  \end{description} } 
\end{thesis}
This is pretty much in the same sense as, by Church's celebrated thesis, a function $f$ is Turing-machine  computable iff $f$ is algorithmically computable according to everyone's reasonable intuition.

It should be also mentioned that, unsurprisingly, brimplicative reducibility turns out to be a conservative generalization of Turing reducibility, commonly accepted in theoretical computer science as the most general relation of reducibility between the traditional, non-interactive sorts of problems.

\section{On intuitionistic logic once again}

According to Kolmogorov's \cite{Kol32} well known thesis, intuitionistic logic is a logic of problems. This thesis was stated by Kolmogorov in rather abstract, philosophical terms. No past attempts to find a strict and adequate mathematical explication of it have fully succeeded. The following theorem tells a partial success story (``partial'' because it is limited to only positive propositional fragment of intuitionistic logic): 

\begin{theorem}\label{blah} [Japaridze \cite{Japapal}; Mezhirov and Vereshchagin \cite{Ver2010}] The positive (negation-free) propositional fragment of Heyting's intuitionistic calculus is sound and complete with respect to the semantics of CoL, with intuitionistic implication understood as $\intimpl$, conjunction as $\adc$ and disjunction as $\add$. 
\end{theorem} 

As for the intuitionistic operators not mentioned in the above theorem, CoL sees the intuitionistic universal quantifier as 
$\ada$, existential quantifier as $\ade$, and negation as what it calls 
{\em brefutation} $\brneg$, defined by 
\[\brneg A  \ =_{def}   A\intimpl \bot.\footnote{As we remember from Section \ref{s3}, the meaning of the logical constant $\bot$ in CoL is standard: this is an always-false proposition, i.e., the elementary game automatically lost by the machine.} \]
So, formula (\ref{e1}) from Section \ref{s2} should in fact have been written as 
\begin{equation}\label{e3}\begin{array}{c}(\brneg P\intimpl A\add B)\adc(\brneg Q\intimpl  C\add D)\adc\brneg(P\adc Q)\intimpl\\
 (\brneg P\intimpl A)\add (\brneg P\intimpl  B) \add (\brneg Q\intimpl C)\add(\brneg Q\intimpl D).\end{array} \end{equation}
This formula, as noted earlier, is valid in CoL but unprovable in Heyting's calculus, making the latter incomplete with respect to the semantics of CoL. At the same time, Heyting's calculus in its full first order language has been shown \cite{int1} to be sound with respect to CoL's semantics. So,  intuitionistic logic --- at least, Heyting's formal version of it --- is a fragment of CoL but, unlike classical logic, ``not quite'' a conservative one. Nevertheless, 
since (\ref{e3}) is the shortest formula known to separate Heyting's calculus from the corresponding fragment of CoL, one can say that Heyting's calculus is quite close to being complete.

\section{Conclusion} 
Computability logic (CoL) is a formal theory of computability in the same sense as classical logic is a formal theory of truth. Its formulas represent computational problems, logical operators stand for operations on such problems,  
and validity means being ``always computable''. Computational problems, in turn, are understood in their most general --- interactive --- sense and, mathematically, are defined as games played by a machine against its environment. 

This article was a brief, informal and incomplete survey of CoL. The latter is not a subject that can be duly introduced within a 1-hour presentation and, in order to well  understand it, one will have to use additional sources. Out of the numerous publications devoted to CoL, the most recommended reading for a beginner are the first ten sections of \cite{Japfin}. An even more comprehensive --- and the most up-to-date --- survey of CoL can be found online in \cite{CL}. 

There was no discussion  of related literature in this article. Such discussions can be found elsewhere, including the already mentioned \cite{Japfin} or \cite{CL}. I just want to point out here that the main precursors of CoL are Lorenzen's \cite{Lor59} dialogue semantics for intuitionistic logic, Hintikka's \cite{Hin73} game-theoretic semantics for classical logic and Blass's \cite{Bla92} game semantics for linear logic, the latter being the closest one.  

The language of CoL is much more expressive than the fragment surveyed in the present article. Important topics not covered here   also include the proof theory of CoL. And, of course, actual and potential applications of CoL outside logic itself. Such applications include theory of (interactive) computation, knowledgebase systems, systems for planning and action, declarative programming languages, constructive applied theories, and more. 

So far the most manifestly realized extra-logical utility of CoL has been  using it as a logical basis for applied theories \cite{clar1}-\cite{clar6}, with  such theories offering substantial advantages over their classical-logic-based counterparts. CoL-based number theory, termed {\em clarithmetic}, will be the subject of a forthcoming paper expected to appear in the next issue of this journal.  

I want to close this article by pointing out that, despite having been evolving for 15 years already, CoL, due to its ambitiousness,  still remains at an early stage of development, with more open questions than answered ones.  A researcher who decides to join the project will find enough interesting material to be occupied with for many years to come. Students are especially encouraged to try.

\end{document}